\documentclass[letterpaper, 10pt, conference]{ieeeconf}
\IEEEoverridecommandlockouts 
\usepackage{amsmath,amscd,amssymb,amsgen,amsfonts,amsbsy}
\usepackage{mathrsfs}
\usepackage[utf8x]{inputenc}
\usepackage{color}
\usepackage{textcomp}
\usepackage{float}
\usepackage{latexsym,graphicx}
\usepackage{tikz}
\usetikzlibrary{arrows.meta,positioning}
\usetikzlibrary{automata,arrows,positioning,calc}
\usepackage{url}
\usepackage{cite}
\usepackage{algorithmic}
\usepackage[ruled,lined]{algorithm2e}
\usepackage{breqn}

\usepackage{subcaption}
\usepackage[caption=false,font=footnotesize]{subfig}

\newcommand{\remove}[1]{}

\newcommand{\comments}[1]{}

\newtheorem{lemma}{Lemma}

\newtheorem{remark}{Remark}
\newtheorem{definition}{Definition}

\usetikzlibrary{shapes.geometric}

\tikzset{
	buffer/.style={
		draw,
		shape border rotate=120,
		isosceles triangle,
		isosceles triangle apex angle=69,
		fill=white,
		node distance=10cm,
		minimum height=10em
	}
}

\makeatother

\title{Outcome-Fair Restless Multi-Armed Bandits for  Stochastic Deadline Scheduling}

\author{
\begin{tabular}{ccc}
  Shakti Sharma &  Rahul Meshram
    \\
    Dept. of Physics,
      & Dept. of Electrical  Engineering, 
   \\
      IIT Madras, Chennai, India.
     &  IIT Madras, Chennai, India.  
 \end{tabular}
}

\begin{document}

\maketitle

\begin{abstract}

We study a restless multi-armed bandit (RMAB) problem for a stochastic deadline scheduling application. RMAB problems are solved using  the Whittle index policy.  The goal in RMAB is to maximize the expected cumulative discounted reward maximization. The Whittle index policy maximizes reward, but is not fair among two classes. In this paper, we introduce  fairness criteria and study an outcome-fair model for RMAB which allows fairness for jobs and users    
structurally disadvantaged demographic classes. 
We formulate an outcome fair stochastic deadline scheduling problem as RMAB, and we develop the outcome fair Whittle index policy. We define a virtual queue mechanism that dynamically enforces long-term completion rate guaranties across demographic groups.
 
We analyze  a standard Whittle index policy and the outcome-fair index policy. We demonstrate the performance of our algorithms with  numerical examples. We compare policies---Whittle index policy (no fairness), input-fairness Whittle index policy, outcome fair Whittle index policy. We observe that the outcome-fair Whittle index policy provides better fairness among classes compared to other policies. We demonstrate a trade off between fairness and profit. This  decreases as the server capacity increases.
\end{abstract}

\section{Introduction}

Restless Multi-Armed Bandit (RMAB) problem is a classical framework in sequential decision-making and planning. At each discrete time step, a decision-maker (the `agent') must choose one of the $N$ available actions (the `arms') to activate. Each arm, when pulled, yields  a reward based on state. The state of all arms evolve at each time step and this evolution is dependent on action like activation or non-activation (passive). 
The goal in RMAB  is the selection of arms subject to budget constraint in each time step to maximize the long term reward function when the state of each arm is evolving according to Markov process. 
 
RMAB problems are known to be PSPACE-hard.
A heuristic  index based policy that assigns a scalar priority score (the index) to each arm at each time step, then activates the $M$ arms with the highest indices. The Whittle index is a popular policy for RMAB problems \cite{Whittle88} and it shown to be near optimal for many resource allocation problems \cite{Weber1990index}.

We consider stochastic deadline scheduling problem  and it has applications to many class of real-world systems, an example electric vehicle (EV) charging station where EVs arrive randomly, each vehicle with a stochastic charging demand (workload) and a hard departure deadline. The station operates $M$ charging ports subject to real-time electricity pricing. When a vehicle departs with an incompletely charged battery, the operator incurs a penalty proportional to the unmet demand. The operator's objective is to maximise total long term infinite horizon discounted profit, combining service revenue, electricity costs, and non-completion penalties.  Other applications of this model include cloud computing (finite-sized jobs with deadlines), patient scheduling and public healthcare intervention, packet scheduling in real-time wireless networks, and data centre workload management.

In this paper, we introduce deadline scheduling problem with fairness constraints.  We propose outcome-fair Whittle index policy. This has advantage in many applications where fairness is essential social welfare criteria. 
Without fairness, a scheduler (decision maker) repeatedly serves the highest-value jobs using Whittle index policy. For example, jobs with long deadlines may never be processed, low-priority users may receive almost no service and some queues may remain permanently backlogged.  In a capacity-constrained system, the scheduler will preferentially activate high-index arms, but these may systematically belong to one demographic group, starving the other.
 Fairness constraints ensure that every job or user eventually receives attention.

\subsection{Related Work}
Restless bandit problem is first introduced by Whittle in \cite{Whittle88} and studied the  Whittle index policy.  RMAB  studied for applications in scheduling problems in wireless networks, 5G networks \cite{Mehta2018rested}, age of information \cite{Hsu2019scheduling} and online caching problem, \cite{Koley2025fresh}.
In  \cite{yu2018deadline}, authors modeled stochastic deadline scheduling as an RMAB, proved indexability, and derived a closed-form Whittle index for the constant-cost case.
In~\cite{raghunathan2008index}, authors demonstrated the Whittle index policy for real-time multicast scheduling in wireless broadcast systems and studied as RMAB, and derived closed form expression for index.  
Recently, RMAB framework is applied for minimizing age of information (AoI) problem, ~\cite{kadota2018scheduling}.

Fairness in RMAB problems  is another direction of  work,\cite{li2022towards,biswas2023fairness, wang2024online, mao2024time} there are no hard deadlines assumed, instead additional fairness constraints introduced and modified version of index policies are developed.
In \cite{ahmed2014fair}, authors examined fair scheduling with deadline guaranties in single-hop networks but did not employ the RMAB framework.

In wireless networks, a packet scheduling problem with hard deadlines has been studied in a work of \cite{hou2009theory}, where authors characterized the feasibility region for timely throughput in single-hop networks and proved the optimality of debt-based policies (Largest debt first policy). Later, this model is extended for heterogeneous real-time traffic over fading wireless channels in \cite{hou2010scheduling}.

The problem of jointly enforcing demographic fairness and hard deadline constraints within an RMAB framework has not been addressed in previous work.
In \cite{kim2014scheduling} studied multicast deadline scheduling in wireless networks and discussed per-class delivery trade-offs, but did not model demographic groups or enforce long-turn completion-rate targets. A work of ~\cite{xie2022architecture} highlighted the need for a fairness-aware management of EV charging infrastructure as a key challenge in future electric energy systems.

\subsection{Our Contributions}

 We study stochastic deadline scheduling problem and this is a class of RMAB problem where queue position is represented as an arm, each queue position is independent of others. The state of each queue position is described using remaining load and time to deadline. The state of each arm evolves independent of other positions, and it changes at each time step.  There is capacity constraints. Thus it is RMAB. We study Whittle index policy. For a simplified model Whittle index formula can be derived. In this policy, the arms with highest indices are scheduled, but these may belong to one class, other class may be starving because arms in these classes are not scheduled. We demonstrated this using a simple example. Our model is extension of \cite{yu2018deadline} with fairness constraints.
 
 We introduce outcome fairness constraints with stochastic deadline scheduling problem. This outcome fairness constraint measures the rate of completion from different classes and maintains the fairness while scheduling. We combine both Whittle index policy and fairness in the index and we propose outcome fairness base Whittle index algorithm. Next we provide computational complexity of this algorithm where we have introduced fairness deficit virtual queue. 
 
 Finally, we provide numerical examples and compare policies---standard Whittle index, input fair index policy and outcome fair index policy, round robin policy. We observe that outcome fair policy have better fairness compare to other schemes.

 Our paper is organized as follows. In Section~\ref{sec:sys-model}, we present system model and preliminary studied on Whittle index policy. We introduce fairness model in Section~\ref{sec:fairRMAB} and  present outcome fair Whittle index policy algorithm and its properties in Section~\ref{sec:outcome-fair-Whittle}. Simulation examples are discussed in Section~\ref{sec:simulation} and concluding remarks in Section~\ref{sec:remark}.

\section{System Model}
\label{sec:sys-model}

We consider a discrete-time stochastic scheduling system with $M$ homogeneous servers and a queue consisting of $N$ job positions, where $N>M$. Time is slotted and indexed by $t\in\{0,1,2,\ldots\}.$  
 The system is modeled as a Markov Decision Process (MDP). At the beginning of each time slot, the scheduler observes the complete system state and allocates at most $M$ servers to the waiting jobs. Each queue position $i\in\{1,\ldots,N\}$ is modeled as an independent restless multi-armed bandit (RMAB) arm. The state of arm $i$ at time $t$ is given by
\begin{equation}
    S_i[t]=\big(B_i[t],\,T_i[t],\,k_i\big),
\end{equation}
which consists of the following components:
\begin{itemize}
    \item $B_i[t]\in\mathbb{Z}_{+}$ denotes the remaining workload of job $i$, such as the remaining charging time of an EV or the remaining computational workload of a task. Whenever a server is assigned to arm $i$, the workload decreases by one unit during that time slot.
    
    \item $T_i[t]\in\mathbb{Z}_{+}$ denotes the remaining time until the job's hard deadline. This quantity decreases by one at every time slot irrespective of the scheduling decision, and the job expires when $T_i[t]=0$.
    
    \item $k_i\in\mathcal{K}$ denotes the demographic or application class associated with job $i$, where $\mathcal{K}$ is a finite set of user groups (e.g., $\mathcal{K}=\{A,B\}$). Different classes may exhibit heterogeneous workload characteristics, deadline distributions, or fairness requirements.
\end{itemize}

Thus, the state of each arm is characterized by three attributes: the remaining workload, the remaining time until deadline, and the associated user class.

In addition to the arm states, the scheduler observes a global exogenous process $c[t]\in\mathcal{C}$ representing the processing cost at time $t$. This cost may capture, for example, the real-time electricity price in an EV charging system or the operating cost of computational resources in a data center, and evolves according to an underlying stochastic process.
A useful state variable is the \emph{laxity} of job $i$ at time $t$, defined as $ L_i[t] = T_i[t] - B_i[t].$
Laxity quantifies the scheduling flexibility of a job. A job with $L_i[t]=0$ must receive service in every remaining time slot to meet its deadline, whereas larger values of $L_i[t]$ indicate greater scheduling flexibility. A job is feasible if and only if $L_i[t]\geq 0$; when $L_i[t]<0$ (equivalently, $B_i[t]>T_i[t]$), it is impossible to complete the job before its deadline under any scheduling policy.

\subsection{Action Space and Capacity Constraint}

At each time slot $t$, the scheduler selects a binary action $a_i[t]\in\{0,1\}$ for each arm $i$, where $a_i[t]=1$ indicates that a server is assigned to job $i$ (active), and $a_i[t]=0$ indicates that the job is not served (passive). Since only $M$ servers are available, the scheduling decisions satisfy $\sum_{i=1}^{N} a_i[t] \leq M,$ $\forall t.$
To cast the problem into the standard RMAB framework, which requires exactly $M$ active arms at every time slot, we augment the system with $M$ dummy arms. Each dummy arm has a fixed state $(0,0)$, yields zero reward, incurs no cost, and remains unaffected by the chosen action. Consequently, the augmented system consists of $N+M$ arms, of which exactly $M$ are activated at every time slot. The original jobs correspond to arms $\{1,\ldots,N\}$, while the dummy arms are indexed by $\{N+1,\ldots,N+M\}$.

\subsection{State Transition Dynamics}

Given the scheduling action $a_i[t]$, the workload evolves according to $B_i[t+1]=\max\{0,\,B_i[t]-a_i[t]\},$ while the remaining time to deadline decreases deterministically as $T_i[t+1]=\max\{0,\,T_i[t]-1\}.$
A job departs the system when either it is completed or its deadline expires. Specifically, if $B_i[t+1]=0$, the job is successfully completed and leaves the system without penalty. If $T_i[t+1]=0$ and $B_i[t+1]>0$, the job departs unfinished and incurs a penalty $F\!\left(B_i[t+1]\right),$ where $F:\mathbb{Z}_+\rightarrow\mathbb{R}_+$ is an increasing convex function satisfying $F(0)=0$. The convexity of $F$ assigns a larger penalty to jobs with greater unfinished workload.
Whenever a job departs, the corresponding queue position is immediately refreshed according to the arrival distribution $Q$. A new job with workload $B$ and deadline $T$ arrives with probability $Q(T,B)$, while the position remains empty with probability $Q(0,0)$.

\subsection{Reward Structure and Objective}

The processing cost is determined by a global stochastic environment state $c[t]\in\mathcal{C}$, which evolves according to a finite-state Markov chain. In contrast, the local state of each arm evolves deterministically given the scheduling action, while job arrivals into vacant queue positions follow the stochastic arrival distribution described earlier. Thus, the overall system dynamics consist of deterministic arm evolution coupled with a stochastic global environment and stochastic job arrivals.

Whenever a server is allocated to job $i$ (i.e., $a_i[t]=1$), the scheduler earns one unit of service reward while incurring a processing cost $c[t]$, resulting in a net reward of $1-c[t]$. Thus, if $B_i[t]>0$ and $T_i[t]>1$, the one-step reward is $(1-c[t])a_i[t]$, and zero otherwise. If the job reaches its final time slot before the deadline, i.e., $T_i[t]=1$, an additional penalty $F(B_i[t]-a_i[t])$ is imposed whenever unfinished workload remains after the scheduling decision. Here, $F:\mathbb{Z}_{+}\rightarrow\mathbb{R}_{+}$ is an increasing convex function satisfying $F(0)=0$, so that larger residual workloads incur a higher penalty. Let $\tilde{S}_i[t]=(S_i[t],c[t])$ denote the extended state of arm $i$, and let $R_{a_i[t]}(\tilde{S}_i[t])$ denote the corresponding one-step reward.

Let $\tilde{S}_i[t]=(S_i[t],c[t])$ denote the extended state of arm $i$. The one-step reward is given by $R_{a_i[t]}(\tilde{S}_i[t])= (1-c[t])\,a_i[t]$ if $B_i[t]>0,\; T_i[t]>1,$  $R_{a_i[t]}(\tilde{S}_i[t])= (1-c[t])\,a_i[t]-F(B_i[t]-a_i[t]),$ if $B_i[t]>0,\; T_i[t]=1,$ and $R_{a_i[t]}(\tilde{S}_i[t])=  =0$ otherwise.
The objective is to determine a scheduling policy $\pi$ that maximizes the expected discounted cumulative reward:
\begin{eqnarray}
V^N(s)
=
\sup_{\pi:\,\sum_{i=1}^{N}a_i[t]\le M,\;\forall t}
\mathbb{E}_{\pi}
\Bigg[
\sum_{t=0}^{\infty}
\beta^t
\sum_{i=1}^{N}
R_{a_i[t]}(\tilde{S}_i[t]) \nonumber \\
\qquad\qquad\qquad\qquad
\Bigm|\,
S[0]=s
\Bigg].
\label{eq:11}
\end{eqnarray}
where $\beta\in(0,1)$ is the discount factor. We  study  deterministic stationary policies.

\subsection{Whittle Index Policy}

Whittle's key idea is to relax the instantaneous resource constraint $\sum_{i=1}^{N} a_i[t] \le M,$ for all $t$ 
by replacing it with a discounted activation constraint and then a Lagrangian relaxation. Introducing a Lagrange multiplier (or subsidy) $\nu\ge0$ for the passive action decouples the original $N$-arm optimization problem into $N$ independent single-arm Markov decision processes. The parameter $\nu$ can be interpreted as a subsidy received whenever an arm is left passive.
For a fixed subsidy $\nu$, the value function of arm $i$ is
\begin{equation}
V_i^\nu(s)
=
\sup_{\pi}
\mathbb{E}_{\pi}
\left[
\sum_{t=0}^{\infty}
\beta^t
R_{a_i[t]}^\nu(\tilde{S}_i[t])
\,\Big|\,
\tilde{S}_i[0]=s
\right],
\label{eq:2}
\end{equation}
where the subsidized one-step reward is
\begin{equation}
R_{a_i[t]}^\nu(\tilde{S}_i[t])
=
R_{a_i[t]}(\tilde{S}_i[t])
+
\nu\,\mathbf{1}\{a_i[t]=0\}.
\label{eq:3}
\end{equation}

The subsidy increases the attractiveness of the passive action. Consequently, as $\nu$ increases, the set of states in which passivity is optimal expands monotonically. This monotonicity is defines the Whittle indexability.
The optimal value function  are as follows. 
\begin{equation}
V_i^\nu(s)
=
\max
\left\{
Q_i^\nu(s,0),
\;
Q_i^\nu(s,1)
\right\},
\end{equation}
where
\begin{align}
Q_i^\nu(s,0)
&=
R_0(s)+\nu
+
\beta
\sum_{s'}
P(s'|s,0)
V_i^\nu(s'),
\label{eq:5}
\\
Q_i^\nu(s,1)
&=
R_1(s)
+
\beta
\sum_{s'}
P(s'|s,1)
V_i^\nu(s').
\end{align}

The optimal action in state $s$ is therefore
\[
a^*(s)
=
\arg\max_{a\in\{0,1\}}
Q_i^\nu(s,a).
\]

\begin{lemma}
  $Q_i^\nu(s,0),$ $Q_i^\nu(s,0),$ and $V_i^\nu(s)$ are non-decreasing piecewise linear and  convex in $\nu$ for fixed $s$  and each $i.$ 
\end{lemma}
The Proof of this lemma is using the mathematical induction technique. We skip the details of the proof due to space constraint. The similar technique is used in \cite[Proof of Lemma $1,$ Appendix A]{Mittal2023indexability}.

\begin{remark}
Note that $R_{a_i[t]}(\tilde{S}_i[t])$ is non-decreasing in $a_i[t]$ and it is independent of state if $B_i[t]>0$ and $T_i[t] >1.$  Reward $R_{a_i[t]}(\tilde{S}_i[t])$ is dependent on state only through penalty for not meeting deadline, i.e.,$B_i[t]>0$ and $T_i[t] =1.$ 
Moreover, jobs arrival is dependent current state $S_i[t].$ 
\end{remark}
Then we have the following lemma. 
\begin{lemma}
    For fixed $\lambda,$ the optimal policy for each arm $i,$ $\pi_i^{\nu}(s)$ is non-decreasing in $s.$ That is, the optimal policy is  a single threshold type in  state $s.$
\end{lemma}
Proof of lemma is skipped due space constraint, The proof is analogous to proof of Theorem $1$ in \cite[Appendix B]{Mittal2023indexability}.

\subsection{Indexability and the Whittle Index}

\begin{definition}[Indexability]
Let
\[
P_i(\nu)=\{s\in S_i:a = 0 \text{ is optimal under subsidy }\nu\}
\]
denote the passive set of arm $i$. Arm $i$ is \emph{indexable} if $P_i(\nu)$ expands monotonically from $\emptyset$ to the entire state space $S_i$ as $\nu$ increases from $-\infty$ to $+\infty$. An RMAB is said to be indexable if every arm is indexable.
\end{definition}

\begin{lemma}
    Each arm is indexable. Further, RMAB is indexable.
\end{lemma}

We  sketch idea  of the proof. Let $\Delta_i^{\nu}(s) = Q_i^\nu(s,0) -  Q_i^\nu(s,1).$ For indexability, we need  to show that $\Delta_i^{\nu}(s)$ is monotone non-decreasing in $\nu$ for each $s.$ This implies that there is  there exists a threshold type policy in $\nu^*(s)$ such that the optimal action $0$ for $\nu \geq \nu^*(s)$ and  the optimal action $1.$ for $\nu < \nu^*(s).$  Define $g(\nu) = V_i^{\nu}(s^{\prime}) - V_{i}^{\nu}(s)$ where $s = (B,T)$ and $s^{\prime} = (B +1 , T),$ observe that $g(\nu)$ is difference of value functions, it is continuous and piecewise linear. Then there exists  $\underline{\nu}$ and $\overline{\nu}$ such that  $\frac{\partial g(\nu)}{ \partial \nu} \geq -1$ for $\nu \in [\underline{\nu}, \overline{\nu}]$ and $\frac{\partial g(\nu)}{ \partial \nu}  = 0$ outside this region. Using this, indexability is proved in \cite[Theorem $1$, Appendix A]{yu2018deadline}. 
\begin{definition}[Whittle Index]
For an indexable arm $i$, the Whittle index at state $s$ is the smallest subsidy that makes the active and passive actions equally desirable:
\begin{eqnarray*}
\nu_i(s)
=
\inf
\left\{
\nu: Q_i^\nu(s,0)
\geq  Q_i^\nu(s,1)
\right\},
\end{eqnarray*}
\end{definition}

For the Whittle index formula, we solve the following equation  for $\nu:$ $Q_i^\nu(s,0) - Q_i^\nu(s,1)=0,$  and index formula is derived. Since the immediate reward structure  is simple for our problem, we can explicitly compute the  index formula. 
However,if  reward structure is complex, then the closed form expression may not be possible. We need to compute the index numerical using value iteration algorithm. The analysis of algorithm of numerical computation of index follows from two-timescale stochastic approximation algorithm. Numerical computation method of index is given in  \cite{Kaza2024constrained,Avrachenkov2026lagrangian}.  With complex transition dynamics and reward structure, it is difficult to  show indexability, in that we can employ heuristic policy like online rollout policy, \cite{Meshram2020}.

In \cite{yu2018deadline}, the indexability is proved and  the Whittle index formula for constant processing cost, i.e., $c[t]=c_0$ is obtained. It is as follows:  $\nu_i(B,T)
=0$ if $B=0.$ $\nu_i(B,T)=1-c_0$ if $1\le B\le T-1.$
$\nu_i(B,T)=1-c_0+\beta^{T-1}  \left[F(B-T+1)-F(B-T)\right]$ if $T\le B.$
The Whittle index of each dummy arm is $\nu_i(0,0,c_0)=0$.
When $B=0$, the job has already been completed and therefore has zero priority. If $1\le B\le T-1$, the job is feasible and can be completed before its deadline, so its index equals the net processing reward $1-c_0$.

\begin{definition}[Whittle Index Policy]
At each decision epoch, compute the Whittle index $\nu_i(\tilde{S}_i[t])$ for every arm $i=1,\ldots,N$, and activate the $M$ arms with the largest indices. Ties, if any, they are broken uniformly at random.
\end{definition}

The Whittle index policy is computationally efficient, requiring $\mathcal{O}(N\log N)$ operations to compute  the indices.

\section{Fairness in RMAB: The Demographic Fairness}
\label{sec:fairRMAB}

\begin{figure}[htbp]
    \centering
\includegraphics[width=0.7\columnwidth]{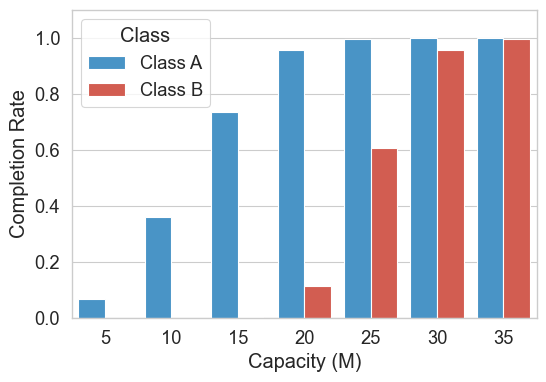}
    \caption{Exp 1: Standard Whittle Inherent Bias. Assuming the probability of arrival of class A user as 70 percent and class B user as 30 percent, we varied the capacity from 20-40 for N=50.}
    \label{Fig-1-Bias-Whittle}
\end{figure}

In Figure~\ref{Fig-1-Bias-Whittle}, we illustrates the demographic bias induced by the standard Whittle index policy. We consider two demographic classes with heterogeneous job characteristics in a queue of size $N=50$. Jobs from the advantaged class (Class A) are generated with workload $B\sim U(2,6)$ and deadline $T\sim U(3,15)$, resulting in relatively high-laxity jobs. In contrast, jobs from the disadvantaged class (Class B) are generated with workload $B\sim U(7,12)$ and deadline $T\sim U(1,5)$, yielding predominantly low-laxity jobs. This example highlights how heterogeneity in job characteristics can lead to systematic differences in scheduling priority under the standard Whittle index policy. Hence unequal service allocation across the two demographic groups. As number of server increases, this bias reduces.

This motivates fairness models which can ensure fairness to different groups. We study two fairness models.  

\subsection{Input Fairness and Outcome Fairness models}

\paragraph{Input Fairness}
Input fairness requires that each demographic class receives a minimum long-run fraction of the available service opportunities. Let $\mathcal I_k=\{i:k_i=k\}.$
denote the set of jobs belonging to class $k$. The input fairness constraint is
\begin{equation}
\liminf_{T\rightarrow\infty}
\frac{1}{T}
\mathbb{E}
\left[
\sum_{t=0}^{T-1}
\sum_{i\in\mathcal I_k}
a_i[t]
\right]
\ge
\eta_k,
\qquad
\forall k\in\mathcal K,
\label{eq:inputfair}
\end{equation}
where $\eta_k$ is the minimum service allocation rate for class $k$.

This notion guarantees that each class receives a prescribed share of server capacity. However, it does not ensure that the allocated service is sufficient to complete jobs before their deadlines. Consequently, a class may satisfy \eqref{eq:inputfair} while still experiencing a high deadline miss rate.

\paragraph{Outcome Fairness}
It requires that each demographic class achieves a prescribed minimum long-run completion rate, thereby accounting for differences in workloads, deadlines, and scheduling decisions. Consequently, outcome fairness provides a stronger  notion of equity in deadline-constrained systems. Motivated by these considerations, we adopt outcome fairness as the fairness criterion throughout this paper.
\begin{definition}[Outcome Fairness]
Let $A_k[t]$ and $C_k[t]$ denote the cumulative numbers of arriving and successfully completed jobs, respectively, for demographic class $k$ up to time $t$. For a prescribed target completion rate $\eta_k\in[0,1]$, the scheduling policy satisfies \emph{outcome fairness} if
\begin{equation}
\liminf_{T\rightarrow\infty}
\frac{C_k[T]}{A_k[T]}
\ge
\eta_k,
\qquad
\forall k\in\mathcal K,
\label{eq:13}
\end{equation}
provided $A_k[T]>0$.
\end{definition}
Constraint \eqref{eq:13} guarantees that, in the long run, at least an $\eta_k$ fraction of jobs from demographic class $k$ are completed before their deadlines. The parameters $\{\eta_k\}_{k\in\mathcal K}$ are specified by the system designer and may differ across demographic groups to account for heterogeneous workloads or to mitigate structural disadvantages.

\section{The Outcome-Fair Whittle Index Policy}
\label{sec:outcome-fair-Whittle}

To enforce the long-run outcome fairness constraint in Eqn.~\eqref{eq:13}, we adopt the Lyapunov virtual queue framework, following the delivery debt approach of  \cite{hou2014scheduling} for deadline-constrained wireless scheduling. The  idea is to associate a virtual queue with each demographic class, whose backlog represents the accumulated fairness deficit. Classes that fall below their target completion rates accumulate larger virtual queues and are consequently assigned higher scheduling priority.

\begin{definition}[Fairness Deficit Virtual Queue]
For each demographic class $k\in\mathcal{K}$, let $\lambda_k[t]\ge0$ denote the fairness deficit virtual queue, initialized as $\lambda_k[0]=0$. The queue evolves according to
\begin{equation}
\lambda_k[t+1]
=
\Big[
\lambda_k[t]
+
\alpha\big(\eta_kA_k[t]-C_k[t]\big)
\Big]^+,
\label{eq:fairness-deficit}
\end{equation}
where $[x]^+=\max\{x,0\}$, $\alpha>0$ is a step-size parameter, $A_k[t]$ and $C_k[t]$ denote the numbers of arrivals and successful completions, respectively, for class $k$ during time slot $t$, and $\eta_k$ is the target completion rate.
\end{definition}
The virtual queue $\lambda_k[t]$ measures the cumulative fairness deficit of demographic class $k$. Whenever the observed completion rate falls below the prescribed target, i.e., $C_k[t]<\eta_kA_k[t]$, the queue increases, thereby assigning higher priority to jobs from class $k$ in future scheduling decisions. Conversely, when the target is exceeded, the queue decreases, reducing the additional priority assigned to that class.

Feasibility Indicator: The feasibility of job $i$ at time $t$ is defined by
$\delta_i[t]=
\mathbf{1}_{\{B_i[t]\le T_i[t]\}},$
where $\delta_i[t]=1$ indicates that the job can still be completed before its deadline, while $\delta_i[t]=0$ indicates that completion is impossible under any future scheduling policy.

\begin{definition}[Outcome-Fair Whittle Index]
The Outcome-Fair Whittle Index of arm $i$ at time $t$ is defined as
\begin{equation}
\nu_i^{\mathrm{fair}}[t]
=
\nu_i(B_i[t],T_i[t],c[t])
+
\lambda_{k_i}[t]\delta_i[t],
\label{eq:16}
\end{equation}
where $\nu_i(\cdot)$ is the baseline Whittle index, $\lambda_{k_i}[t]$ is the virtual queue associated with the demographic class of job $i$, and $\delta_i[t]$ is the feasibility indicator.
\end{definition}

The first term represents the economic priority of the job under the standard Whittle policy, while the second term provides an adaptive fairness incentive. Consequently, jobs belonging to demographic classes with larger accumulated fairness deficits receive higher scheduling priority, but only if they remain feasible. 
 At each time slot, the scheduler computes the outcome-Fair Whittle index for every active job, activates the $M$ jobs with the largest indices, and updates the fairness virtual queues using the observed arrivals and completions.

 The sufficient conditions on fairness-deficit virtual queue in Eqn~\eqref{eq:fairness-deficit} for feasibility optimality  is analyzed using the Lyapunov drift criteria, see \cite[Theorem $2$]{hou2010scheduling}. 
 Outcome fair Whittle index policy schedules $M$ arms with highest indices $\{\nu_i^{fair}\},$
 and this ensures that each class is scheduled and fairness criteria with completion rate is maintained.
 If it is not maintained for arm $i$ in class $k$, then $\lambda_{k_i}[t]$ increases for arm $i,$ it increases fair index for arm $i$ and it is scheduled to the server. The fair index policy balances the tradeoff between total expected reward and fairness.  
The computational complexity of outcome fair Whittle index policy at each decision epoch is $O(N\log N).$

\section{Simulation Results}
\label{sec:simulation}

We present numerical experiments comparing three scheduling policies: the standard Whittle index policy, the Input-Fair Whittle index policy, and  Outcome-Fair Whittle index policy. The standard Whittle index policy schedules jobs solely according to their Whittle indices, computed using the closed-form index expression. The Input-Fair Whittle policy enforces a fixed per-class server quota but does not adapt to differences in workload or deadline characteristics across classes. In contrast, the proposed Outcome-Fair Whittle policy dynamically adjusts scheduling decisions through a virtual queue to achieve desired long-term completion-rate targets.

\begin{figure}[htbp]
    \centering
\includegraphics[width=0.85\columnwidth]{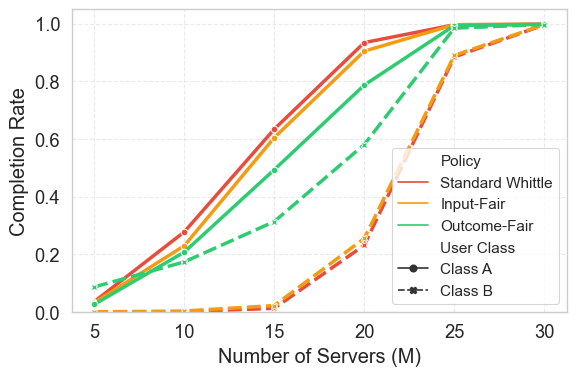}
    \caption{Completion rates vs Server capacity for a simple two class case, illustrating fair completion rates for outcome-fair whittle policy.}
    \label{fig:completion-rate-1}
\end{figure}

In the simulation example~1 (Figure~\ref{fig:completion-rate-1}), we consider $N=50$ parking spots and cost of fixed $c_0 =0.2.$ discount factor $\beta=0.99$, and terminal penalty $F(B)=1.5 B$. Class~A users ($70\%$ of arrivals) generate low-demand, relaxed-deadline jobs with $B\sim U(2,6)$ and $T=B+U(3,12)$, while Class~B users ($30\%$ of arrivals) generate high-demand, tight-deadline jobs with $B\sim U(6,13)$ and $T=B+U(1,5)$. Each vacant parking spot receives a new arrival with probability $0.9.$ The number of available servers is varied over $M\in{5, 10, 15,20,25,30}$, while all other parameters remain fixed. Each policy is simulated over 5000 time slots. The Input-Fair Whittle policy reserves $30\%$ of the available servers for Class~B users. The proposed Outcome-Fair Whittle policy instead employs a virtual queue with target completion rate $\eta_B=0.5$ and step size $\alpha=5\times10^{-5}.$ Figure~\ref{fig:completion-rate-1} present the completion rates of Classes~A and B as functions of server capacity. Figure illustrate that outcome-fair Whittle index policy consistently achieves a higher completion rate for the disadvantaged Class~B users than the input-fair Whittle policy. Observe that in the figure  solid lines for class A and dotted lines for class B. As server capacity increase, the completion rate increases  for both classes. 

\begin{figure}[htbp]
    \centering

    \begin{subfigure}[t]{0.45\columnwidth}
        \centering
        \includegraphics[width=\linewidth]{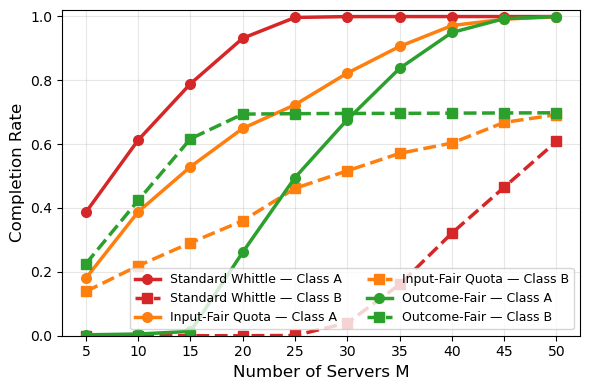}
        \caption{$N=80$ Completion rate vs server capacity}
        \label{fig:completion-rate-2}
    \end{subfigure}
    \hfill
    \begin{subfigure}[t]{0.45\columnwidth}
        \centering
        \includegraphics[width=\linewidth]{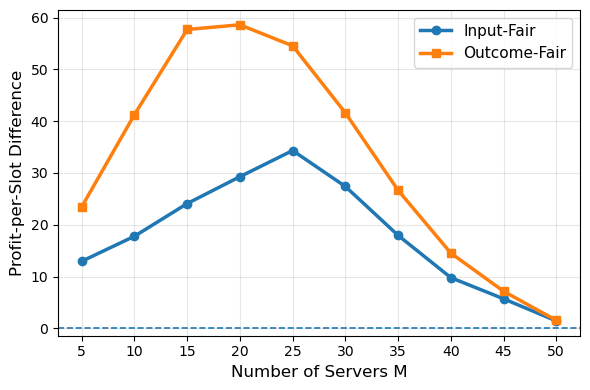}
        \caption{$N=80$ Differences in profit}
        \label{fig:loss-1}
    \end{subfigure}

    \caption{Comparison of scheduling policies as the number of servers increases. Figure shows the completion rates for the two user classes and difference in profit.}
    \label{fig:completion-rate}
\end{figure}

In example 2 (Figure~\ref{fig:completion-rate-2}), we illustrate the completion rate vs server capacity, outcome fair policy increases the fairness. However, there is  decrease in the total profit. There is trade-off between fairness and profit.

In example 2, Figure~\ref{fig:loss-1}, we compare difference of profit from  1) standard whittle index policy and input fair policy, it is denoted a blue line. 2)  Standard whittle index policy and outcome fair policy, it is denoted by an orange line. As the number of  servers increases, this loss in profit difference decreases. More details on simulations of example~2 are given in Appendix. 

\section{Concluding Remarks}
\label{sec:remark}
We studied outcome fairness for restless multi-armed bandits for deadline scheduling problem, where we proposed the outcome fair Whittle index policy. We presented analysis on the Whittle index policy and it extension to outcome fair index policy. 

We illustrated numerical examples  and it demonstrate performance  of outcome fair Whittle index policy and it is compared with standard Whittle index policy and input fair index policy. We observed that there is a trade-off between profit and fairness. The completion rate is improved in outcome-fair index policy while profit difference between standard Whittle index and outcome fair index policy  is high. 

In the future we plan to investigate further on fairness vs profit trade-off for  deadline scheduling problems with varying immediate reward structures.

\bibliographystyle{IEEE}

\bibliography{bandits-ref}

\appendix

\subsection{Study of Example 2: General Model for Deadline Scheduling}

At each time step, the scheduler selects at most $M$ jobs to receive one unit of service. The objective is to maximize the cumulative discounted  reward over the infinite horizon.

We now modify the reward structure and include immediate service reward,
 completion bonus, deadline failure penalty, and wasted service penalty.
Suppose 
\begin{itemize}
    \item $a_i[t]\in\{0,1\}$ denote whether job $j$ is served at time $t$,
    \item $k_i\in\{\text{A},\text{B}\}$ denote the class of job $j$,
    \item $c_t$ denote the electricity price at time $t$,
    \item $v_k$ denote the service value of class $k$,
    \item $G_k$ denote the completion bonus,
    \item $F_k$ denote the deadline penalty coefficient,
    \item $w_k$ denote the wasted-service penalty coefficient.
\end{itemize}

Whenever a selected job receives one unit of processing, its remaining workload decreases by one unit,
\[
B_i[t+1]=B_i[t]-1,
\]
and the scheduler earns the immediate reward

\[
r_{\mathrm{service}}^{(i)}[t]
=
v_{k_i}-c_t.
\]
Here, $v_{k_i}$ represents the economic value of serving one unit of workload, while $c_t$ denotes the electricity price during slot $t$. Consequently, serving jobs during periods of low electricity prices yields higher profit.

The total immediate reward collected at time $t$ is 
\[
R_{\mathrm{service}}[t]
=
\sum_{i\in\mathcal{Q}[t]}
a_i[t]
\left(
v_{k_i}-c_t
\right),
\]
where $\mathcal{Q}[t]$ denotes the set of jobs currently present in the system. A job is successfully completed whenever $B_i[t]=0.$
Successful completion yields an additional terminal reward $r_{\mathrm{completion}}^{(i)}
=
G_{k_i},$
where $G_{k_i}$ depends on the job class.
The total completion reward at slot $t$ is
\[
R_{\mathrm{completion}}[t]
=
\sum_{i\in\mathcal{C}[t]}
G_{k_i},
\]
where $\mathcal{C}[t]$ denotes the set of jobs completing during slot $t.$

 \subsubsection{Deadline Failure Penalty}

If the job reaches its final time slot before the deadline, i.e., $T_i[t]=1$, an additional penalty $F(B_i[t]-a_i[t])$ is imposed whenever unfinished workload remains after the scheduling decision. Here, $F:\mathbb{Z}_{+}\rightarrow\mathbb{R}_{+}$ is an increasing function satisfying $F(0)=0.$ For simplicity,  we assume that $F(B_i[t]) = P_i B_i[t],$ where  $P_i$ is penalty and $B_i[t] > 0.$

Thus, reward to 
the scheduler for the remaining unfinished workload is 
\[
r_{\mathrm{remaining}}^{(i)}[t]
=
-
P_{k_i}
B_i[t].
\]

The corresponding total penalty is

\[
R_{\mathrm{remaining}}[t]
=
-
\sum_{i\in\mathcal{F}[t]}
P_{k_i}
B_i[t],
\]
where $\mathcal{F}[t]$ denotes the set of jobs leaving unsuccessfully.

\subsubsection{Wasted Service Penalty}

Suppose an unfinished job has already received $E_i[t]$ units of service before missing its deadline.
Since these resources do not produce a completed job, the scheduler incurs the additional penalty
\[
r_{\mathrm{waste}}^{(i)}
=
-
z_{k_i}
E_i[t].
\]
The total wasted-service penalty equals
\[
R_{\mathrm{waste}}[t]
=
-
\sum_{i\in\mathcal{F}[t]}
q_{k_i}
E_i[t].
\]

This term discourages allocating resources to jobs that are unlikely to finish before their deadlines.

Combining all four components, the one-step reward is
\[
\begin{aligned}
R[t]
=&
\sum_{i\in\mathcal{Q}[t]}
a_i[t]
\left(
v_{k_i}-c_t
\right)
+
\sum_{i\in\mathcal{C}[t]}
G_{k_i}
\\
&
-
\sum_{i\in\mathcal{F}[t]}
\left(
P_{k_i}B_i[t]
+
z_{k_i}E_i[t]
\right).
\end{aligned}
\]

The reward structure simultaneously captures the economic value of service, the cost of electricity, the incentive to complete jobs, and the opportunity cost of wasting limited charging or processing resources on jobs that eventually miss their deadlines.

We have the heuristic  index  $W_i(B,T,k).$

\[
\begin{aligned}
\widetilde W_k(B,T)
=&\;
v_{k_i}-c
\\
&
+\mathbf 1_{\{B=1\}}G_{k_i}
\\
&
+\mathbf 1_{\{B\ge T\}}
\beta^{T-1}
\left(
P_{k_i}
+
z_{k_i} E_i
\right),
\end{aligned}
\]

The first term $v_{k_i}-c$
represents the net profit obtained by serving the job during the current time slot. Jobs with larger service value or lower electricity prices receive higher priority.
The second term $\mathbf1_{\{B=1\}}G_{k_i}$  is completion incentive, it is activated whenever one additional unit of service completes the job. 
The final term $\mathbf1_{\{B\ge T\}}
\beta^{T-1}
(P_{k_i}+z_{k_i}E_i)$ 
is active only for critical jobs, i.e., jobs whose remaining workload is at least as large as the remaining time before departure. This term estimates the discounted future loss that can be avoided by serving the job immediately.
The quantity $P_{k_i}+q_{k_i}E_i$
represents the total expected terminal loss.

\subsubsection{Scheduling Policy}

At every decision epoch, the heuristic index is computed for every job in the queue. The scheduler then selects the $M$ jobs with the largest values.

\[
\mathcal A(t)
=
\arg\max_{|\mathcal S|=M}
\sum_{j\in\mathcal S}
\widetilde W_j(B,T),
\]

This policy preserves the computational efficiency of Whittle-index scheduling while incorporating the richer economic reward structure of the proposed charging model.

\subsubsection{Discussion}

The proposed score should be viewed as a \emph{Whittle-inspired heuristic} rather than an exact Whittle index. An exact Whittle index would require solving the single-job subsidized dynamic programming problem and determining the subsidy for which the active and passive actions are equally valuable,

\[
Q^\nu(B,T,1)=Q^\nu(B,T,0).
\]
The resulting index generally does not admit a closed-form expression under the proposed reward model because of the interaction between completion bonuses, heterogeneous service values, and class-dependent terminal penalties. Nevertheless, the heuristic captures the dominant economic trade-offs while maintaining the low computational complexity characteristic of index policies.

\subsubsection{Details on Simulation}
In this example, we demonstrate with this general reward, the completion rate vs number of servers capacity. 
There are two classes are considered.
The outcome fair algorithm improves the completion rate for class $B$ compared to a heuristic Whittle index policy. This is illustrated in the Figure~\ref{fig:completion-rate-2}. As number of server increases, the completion rate improves. The completion rate of outcome fair index policy and Whittle index policy matches when the server capacity reaches to $45.$

We also plot  the profit difference between Whittle index policy and outcome fair index policy, profit difference Whittle index policy and input fair index policy.  It illustrates the trade-off between fairness and profit. It is given in Figure~\ref{fig:loss-1}. Interestingly, the profit difference is high when server capacity  at 20-25, while profit difference is minimum, reaches to $0$ when server capacity increases  $50.$

The following parameters are used in simulations. $N=80$ parking spots and cost is $0.2,$ $\beta = 0.99$ 
Class~A users ($70\%$ of arrivals) generate low-demand, relaxed-deadline jobs with $B\sim U(2,6)$ and $T=B+U(3,8)$, while Class~B users ($30\%$ of arrivals) generate high-demand, tight-deadline jobs with $B\sim U(8,15)$ and $T=B+U(1,4)$. Service value for class A $v_{A} = 2.2$ and for class B $v_B = 0.9.$ The completion bonus $G_A = 7$ for class A and $G_B = 1.5$ for class B. Failure penalty $P_A = 1.5$ and $P_B = 0.6$ for class A and class B. Waste cost for class A and class B is $E_A = 0.2$ and $E_B = 0.6.$ 
Each vacant parking spot receives a new arrival with probability $0.55.$ The number of available servers is varied over $M\in{5, 10, 15,20,25,30, 35, 40, 45, 50}$, while all other parameters remain fixed. Each policy is simulated over $5000$ time slots. The Input-Fair Whittle policy reserves $30\%$ of the available servers for Class~B users.

\end{document}